# A Framework for Real Time Hardware in the loop Simulation for Control Design

Abdalla O.M., S.A. Hammad, A.H.Yousef
abdalla.osama@gmail.com, sherif_hammad@mentor.com, ahassan@topit.com.eg
1 Sarayat street, Abbasya
Computers and Systems Engineering Department, Ain Shams University, Egypt

***Abstract-***

**This paper presents a simple framework of low cost Kit which can be used in control education and training courses to support hardware in the loop simulation. The kit shows the student or control engineer the effect of delays, noise, and saturation on the control system. The framework is generic and flexible to give the user the ability to test and simulate any controller on any process. The framework uses Matlab® environment which gives the user many tools to build his/her system in a fast and accurate way. Some test cases are presented for using the framework on different controllers.**

## I. Introduction

Embedded systems are designed to control complex plants such as land vehicles, satellites, spacecrafts, Unmanned Aerial Vehicles (UAVs), aircraft, weapon systems, marine vehicles, and jet engines [1-6]. They generally require a high level of complexity within the embedded system to manage the complexity of the plant under control and surrounding environment [3]. These systems may need advanced and intelligent control algorithms and online optimization methods that satisfy robustness and excellent dynamic characteristics [7].

Over the past years, control engineers used simulation extensively as a start point for control design. However, they reported that simulation results and environment are far beyond real time results and environment. The aim of this paper is to describe a framework that enables control engineers to have reality in their simulations. The effects of delays, noise, saturation and digital to analog nonlinearity are incorporated in the simulation environment.

Hardware-in-the-Loop (HIL) simulation is a technique that is used increasingly in the development and test of complex real-time embedded systems [3, 8]. The purpose of HIL simulation is to provide an effective platform for developing and testing real-time embedded systems. HIL simulation provides an effective platform by adding the complexity of the plant under control to the test platform. The complexity of the plant under control is included in test and development by adding a mathematical representation of all related dynamic systems. This technique is very useful in rapid prototyping. These mathematical representations are referred to as the “plant simulation.”

An HIL simulation must also include electrical emulation of sensors and actuators. These electrical emulations act as the interface between the plant simulation and the embedded system under test. The value of each electrically emulated sensor is controlled by the plant simulation and is read by the embedded system under test. Likewise, the embedded system under test implements its control algorithms by outputting actuator control signals. Changes in the control signals result in changes to variable values in the plant simulation.

In some situations, HIL simulation is more efficient than connecting the embedded system to the real plant. Selecting the best model is typically a formula that includes the cost, development duration and safety factors.

The tight development schedules associated with most new automotive, aerospace and defense programs do not allow embedded system testing to wait for a prototype to be available. In fact, most new development schedules assume that HIL simulation will be used in parallel with the development of the plant.

In many cases, the plant is more expensive than a high fidelity, real-time simulator and therefore has a higher-burden rate. Therefore, it is more economical to develop and test while connected to an HIL simulator than the real plant.

HIL simulation is a key step in the process of developing human factors, a method of ensuring usability and system consistency in addition to human-factors research and design. For real-time technology, human-factors development is the task of collecting usability data from man-in-the-loop testing for components that will have a human interface.

In the case of fly-by-wire flight controls development, HIL simulation is used to simulate human factors. The flight simulator includes plant simulations of aerodynamics, engine thrust, environmental conditions, flight control dynamics and more. Prototype fly-by-wire flight controls are connected to the simulator and test pilots evaluate flight performance given various algorithm parameters.

The dynamic capability of the test system must exceed the dynamics of the fastest component within the process under test. This criterion is even more critical when a simulated component of the process is replaced by hardware and Hardware-in-Loop (HiL) methodology is employed [5].

This paper is organized as follow: The problem statement will be described in section II. The proposed framework and its components and different operation modes will be presented in section III. In the same section, real time problems and the proposed solution to overcome and fix these problems will also be described. Some test cases on the proposed framework will be shown in Section IV. The test cases include the use of different controllers like PID and RST with different modes.

## II. Problem Statment

Traditional Control Education and Training usually start with a mathematical model of the industrial process, sensors and actuators. However, senior control engineer in practical industrial systems don't recommend that for junior engineers because there is a large gap between practical system and simulations tools.

For example, some intelligent control algorithms like NARMA-L2 control and identification scheme give positive results in simulation. The same controllers may fail in reality because of the ignorance of the limiting value of control action, the existence of noise or nonlinearity.

In this framework, the aim is to decrease the gap between simulation environment and real time environment by incorporating the effects of the factors mentioned above.

## III. The Proposed Framework

### A. Overview

The framework design goals were simplicity, extensibility and affordability. Simplicity and affordability was ensured by the hardware components chosen. Its hardware consists of low price ADDA card connected to the PC via USB cable. The extensibility is achieved by the selected control design software. It is chosen to be Matlab® Software along with its rich set of Simulink and other toolboxes. The software is a part of the framework to drive such ADDA card. This extensibility feature solved for the "unavailability of control algorithms that incorporate many of the advanced features in the current generation of simulation packages" that is reported in [1].

Although the design is simple, the control engineer can leverage the full functionality of Matlab® and Simulink® to build huge number of different control systems. The ADDA card gives him the ability to validate the proposed control strategy on real time, any process and controller can be simulated and tested.

### B. K8055 ADDA card

Currently all supported ADDA cards in Matlab® are highly expensive for many students and researchers. So, they can not be used in Education nor training for novice control engineers. The K8055 card is chosen in this work because of its low cost price and market availability. Figure 1 compares the prices of typical ADDA cards in the market.

The family of low cost cards lacks high speed sampling and linearity. These drawbacks were overcome successfully by implementing new Matlab® toolbox components.

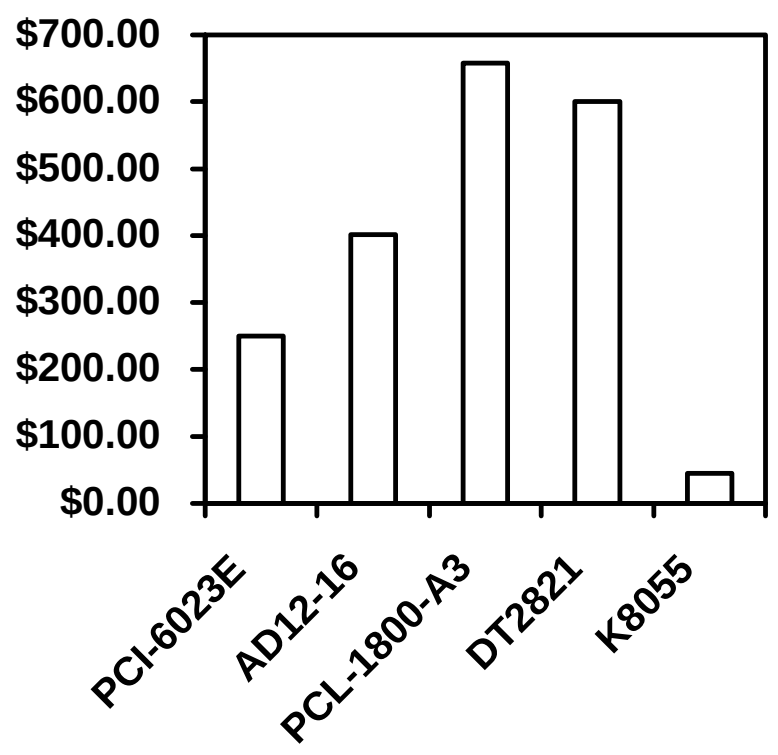

Fig 1. Comparison between typical ADDA Cards prices

The K8055 interface board has 5 digital input channels and 8 digital output channels. In addition, there are two analogue inputs and two analogue outputs with 8 bit resolution. The number of inputs/outputs can be further expanded by connecting more (up to a maximum of four) cards to the PC's USB connectors. All communication software routines are contained in a Dynamic Link Library (DLL) file.

The framework can run on two modes, real time simulation mode and hardware in the loop simulation mode. Fig 2 and Fig.3 show each of these modes respectively. The remaining part of this section will focus on describing these modes.

Testing of the chosen ADDA card as a sample of the non-expensive cards showed some problems. These problems would waste time of the control engineer to be fixed. These problems could be summarized in the following points.

***Non predictable delay:*** the time needed from generating the software signal until it appears steadily on the hardware is called the delay. It's shown in fig 4. After recording the delay in many tests, it was found that the average delay is about 5 cycles. Delay value may vary due to the overall performance of PC.

***The nonstandard operating voltage range***: It was specified that the operating voltage range is from 0.0v to 5.0v. After testing, it was found to be from 0.0v to 4.5v.

***The nonlinearity problem:*** For low cost ADDA cards, the card may suffer from nonlinearity problem in the digital to analog conversion part.

### C. Realtime ToolBox and Simulation modes

Matlab® Toolbox consists of three pairs of blocks. One pair is called DAADx and is used in Real time simulation mode. Two pairs of other blocks named ADCx and DACx are used in Hardware in the loop simulation mode.

To achieve real time simulation, a third party block was added to the Simulink® model called RT Block. The block has been realized using an S-function written in C++ language.

The Real Time blockset is based on the simple concept that, to make the Simulink run with a real-time temporization, the cycle time (the time that Simulink needs to calculate a simulation step, that is function of the hardware and the OS in which Simulink is running) should be lower then the desired simulation step. If this assumption is not valid, no real-time simulation is possible whichever the applied scheduling method is [14].

### D. Real time simulation

In this mode, the process and controller will be simulated on Matlab®. Any system can be simulated in real time by adding the RT block. This mode is represented in figure 2.

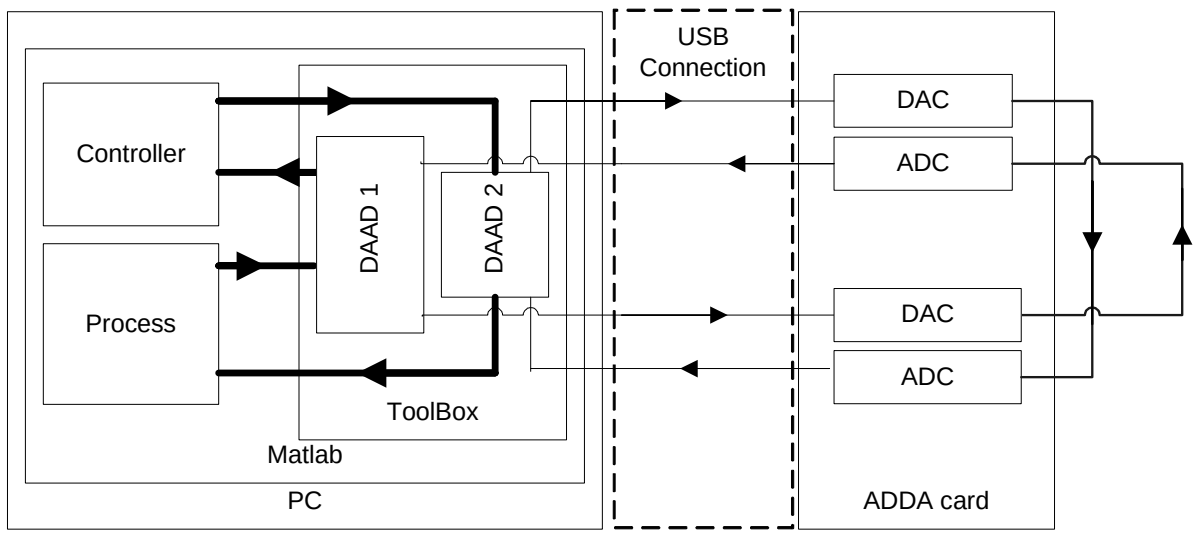

Fig. 2 Real time simulation diagram

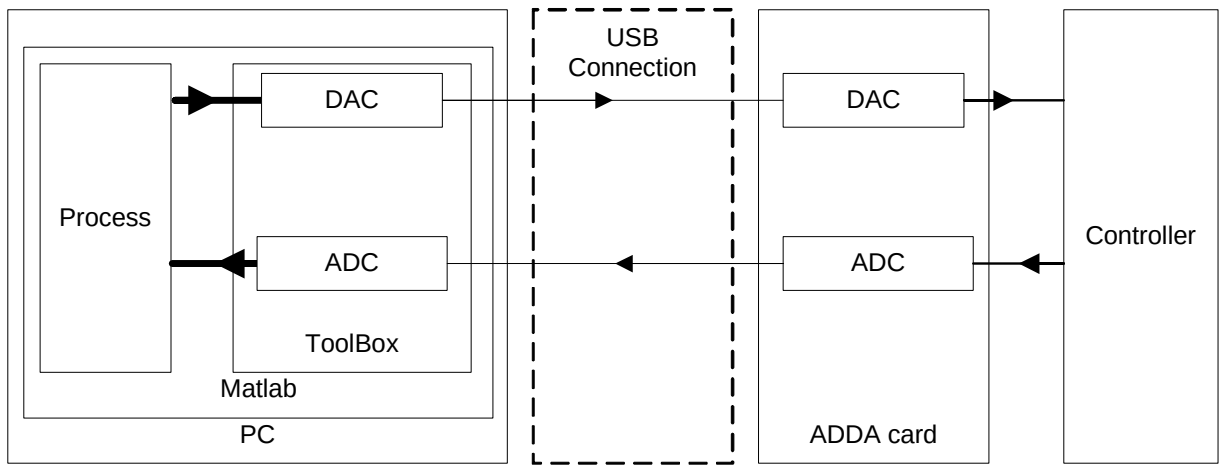

*Fig. 3. Hardware in the Loop Simulation Diagram.*

Figure 2 showed a complete simple closed loop control system that consists of both the process and controller simulated in Matlab® environment, the toolbox DAAD component is used to send the signal via USB connection. The output of Digital to analog converter of the card is hardwired to the input of the analog to digital converter. This setup ensures that the signal is propagated through the card. This ensures also that the control action is limited by the card maximum and minimum values to ensure the practical realization of the controller.

### *E. Hardware in the loop real time simulation*

In this mode the Matlab simulate process only and a real time controller is used. A block diagram of this mode is shown in fig.3.

Figure 3 showed a simple complete closed loop control system. User can interface any controller with the ADDA card and the framework. The MC9S12E128 microcontroller kit was chosen for the all test cases.

The Microcontroller Programming could be simulated by writing the algorithm as an S function in C Code and execute simulation in the simulator environment. This is repeated many times to verify the design and to tune the controller algorithm. This scheme allows rapid testing of the algorithm without need of microcontroller programming.

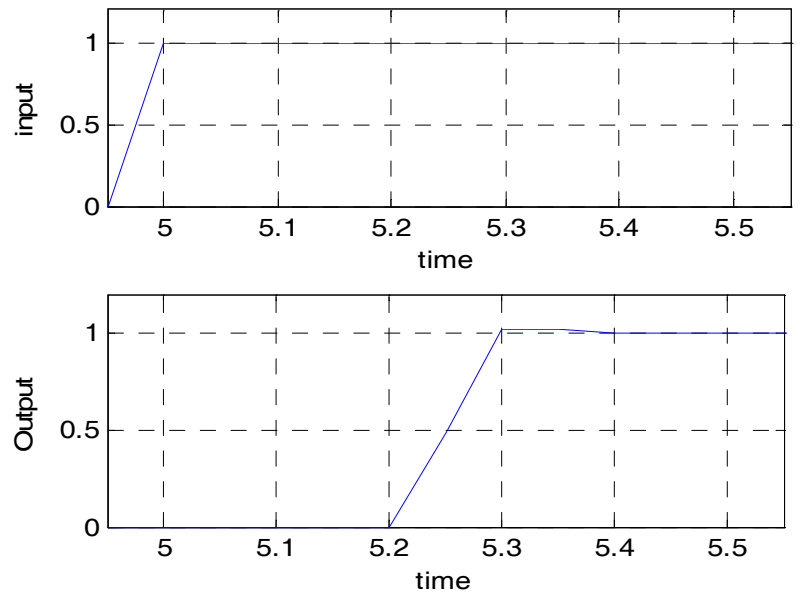

Fig.4. Delay graph.

After having good results, the final microcontroller program will be applied to the PROM of the microcontroller.

### *F. Analog and Digital controller sampling modes*

The sampling time should be decreased to minimize discretization effect. The sample time could be decreased until it reaches Matlab simulation step. In the case of analog controller design, the component sample rate is adjusted automatically to be equal to Matlab step and doesn't need for sample rate setting.

In the case of digital controller design, the user can adjust the requested sample rate for each component depending on the real time constant of each component (Sensor, Actuator,…etc)

### *G. Toolbox implementation*

Toolbox components can be implemented using Matlab s-functions and can be written in MATLAB, C, C++, Ada, or Fortran. It was chosen to write the s-functions in C code and compile it to DLL file. Many problems that were discovered in that card were solved by software solutions.

For example, the delay problem which results in differences between the output of a solution in simulation mode and in real time simulation was solved as follows. Each block has in and out ports. The output signal of the component is known and equal to the input signal. The component software is redesigned to check on the output. If it is not approximately equal to the input value, the value of ADC is read again. An example is shown in fig.4. Matlab will not increment simulation time till ADC reads the correct values which must be within acceptable percentage of error. In this implementation, it was be 2%, this solution increased the load on the PC and so real-time may be lost due to that delay but this problem appears only if the delay time is more than *simulation step – calculation time*. This condition can be avoided in all test cases by selecting larger simulation step.

In case of HIL simulation, delay can't be avoided because it is inherent in the ADDA Card. The control engineer or designer must account for the delay to increase the designed controller robustness.

The second problem of low cost ADDA cards was nonlinearity and saturation. It was expected that the card output will span from 0.0 volts to 5.0 volts. In reality, the range was from 0.0 to 4.5 volts

A software solution that uses both multiplication and lookup table is used to linearize the card and increase its range. This solution compensated the deviation of the signal to the correct value. It's clear that this solution has a drawback on the resolution value of the card which will be decreased slightly.

## IV. TEST CASES ON THAT FRAMEWORK

Although the card has many sides of weakness points due to its low cost, the solution and techniques mentioned above increased its range of applications. The framework is proved to be suitable for many researches.

Many test cases have been implemented successfully on the framework. Different analog and digital control systems with different processes are simulated and implemented in real time. In this section, some cases will be presented.

### *A. Analog PID control on a heat exchanger process*

A real time simulation of a system that consists of an Analog PID and heat exchanger process is implemented. The entire system is shown in fig 5. The error signal is passed to the PID Controller. The control action signal is passed through channel one to the process. The process feedback is collected using the second channel. The response of the system and the control action are shown in fig.6.

The same system is simulated on Matlab Simulink only, in both cases the Simulation time step is 45ms with unit step that starts at 1 sec. Fig 7 shows the control action of the PID and system output

The results reported in fig. 6 and 7 showed similar graphs with the existence of little noise in real time implementation.

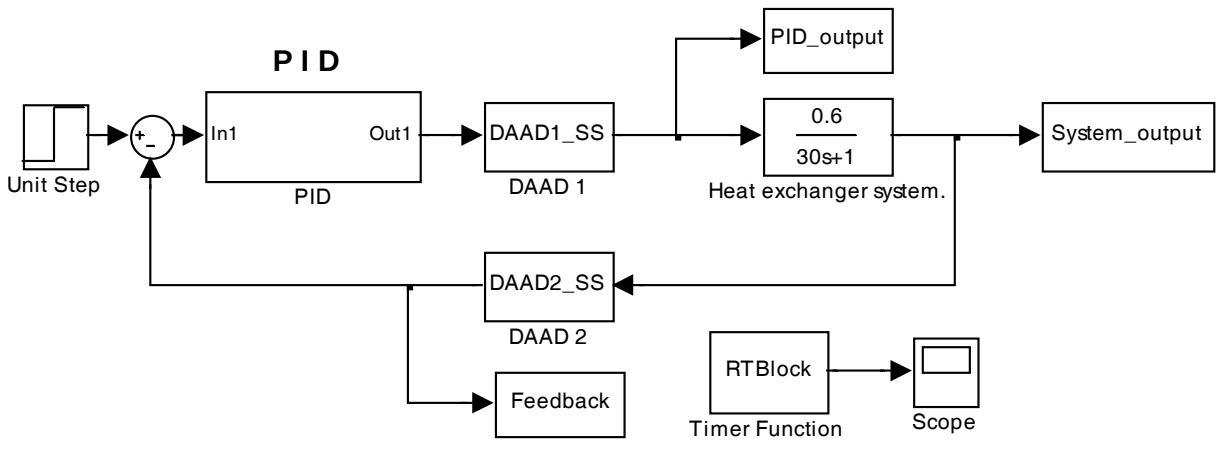

*Fig. 5. Analog PID Controller for Heat Exchanger System*

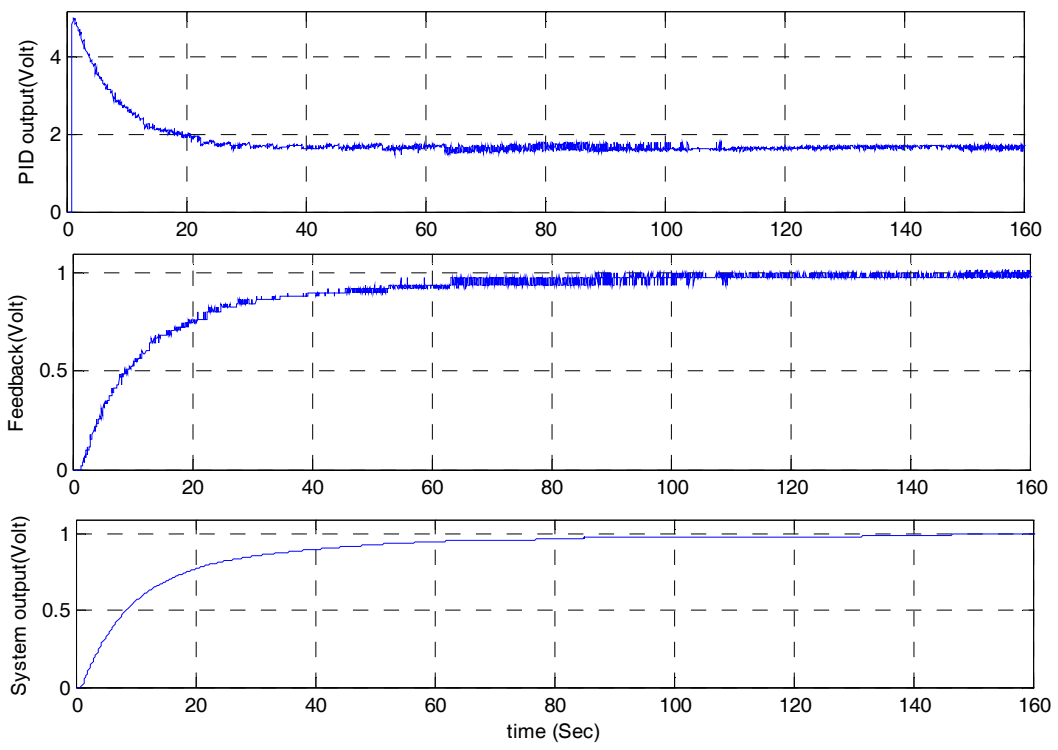

*Fig.6 Framework Result of the PID controlled heat exchanger process*

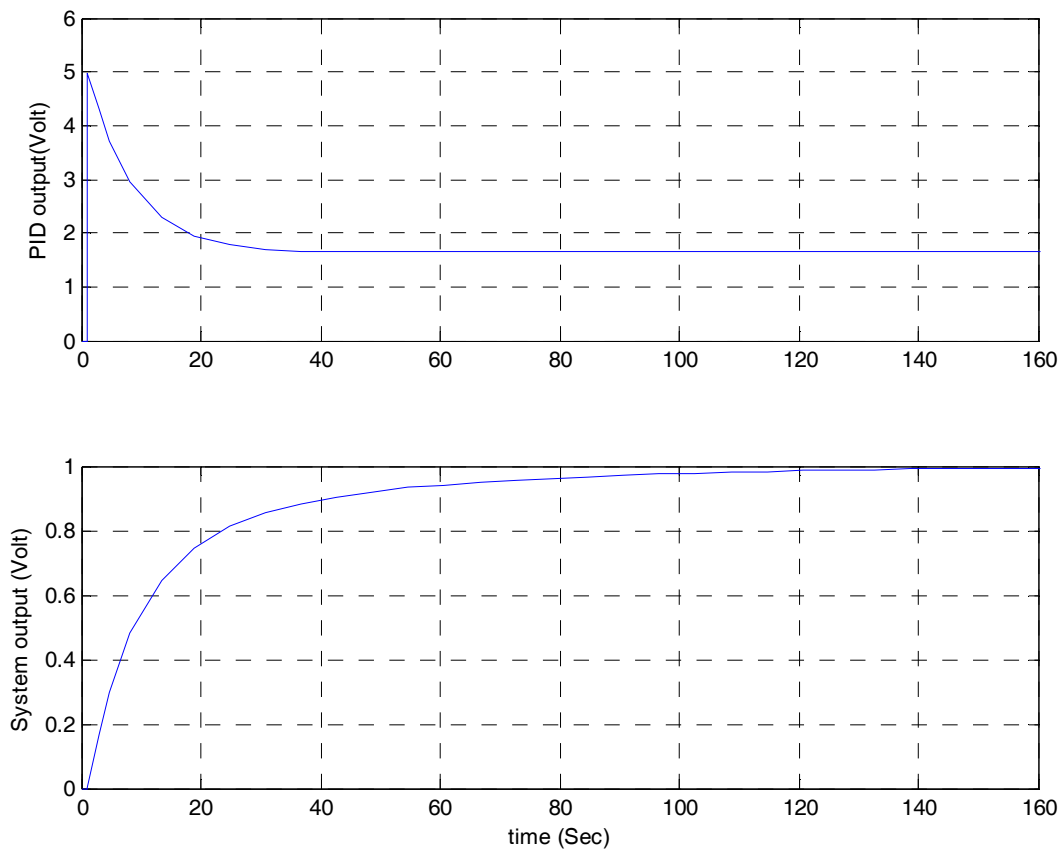

Fig. 7 Matlab simulation result for PID controlled heat exchanger process

### *B. Real Time Simulation for a Digital RST controller on a heat exchanger process*

In [13], an internal mode controller (IMC) that depends on pole placement is proposed. The controller is enhanced to support independent tracking and regulation dynamics.

The RST Controller polynomials are usually selected with these equations.

$$A(q^{-1})\, S(q^{-1}) + q^{-d} B(q^{-1})\, R(q^{-1}) = P(q^{-1}) \qquad (1)$$

$$T(q^{-1}) = A(q^{-1})\, P(q^{-1}) / B(1) \qquad (2)$$

The polynomials B and A are the nominator and denominator polynomials of the process. The total delay of the process is d. The delay element is represented by q. P is the polynomial of the required system dynamics. R,S and T are the polynomials representing the RST Controller.

The equations are solved and the system has been implemented on the framework. It's shown in fig 11. The system is simulated also to compare results obtained from the hardware in the loop simulation mode and normal simulation mode. In both cases, the unit step starts at 0 sec, all controller components discretized with a 1 second sample rate.

As shown in fig 8, the heat exchanger model was simulated in S domain form (continuous) in real time simulation mode. The other model is for the R-S-T digital controller and is represented in Z domain. To enable Matlab to simulate blocks with different rates in the same model, the rate transition block is used. The Rate Transition block transfers data from the output of a block operating at one rate to the input of another block operating at a different rate. The results of Matlab simulation and framework real time simulation are shown in fig 9 and fig 10 respectively.

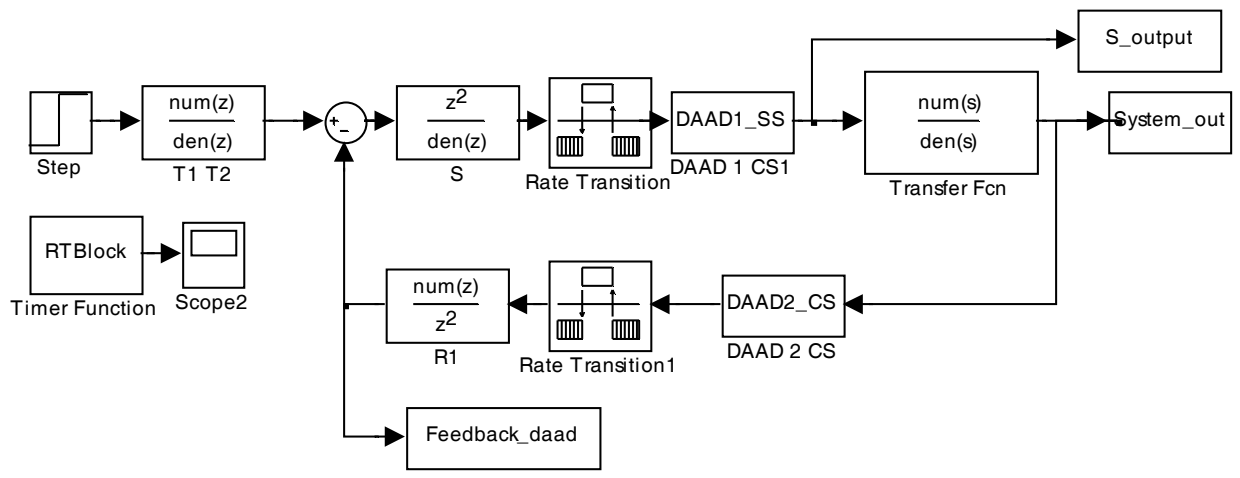

*Fig. 8. RST Controller with a Heat Exchanger System*

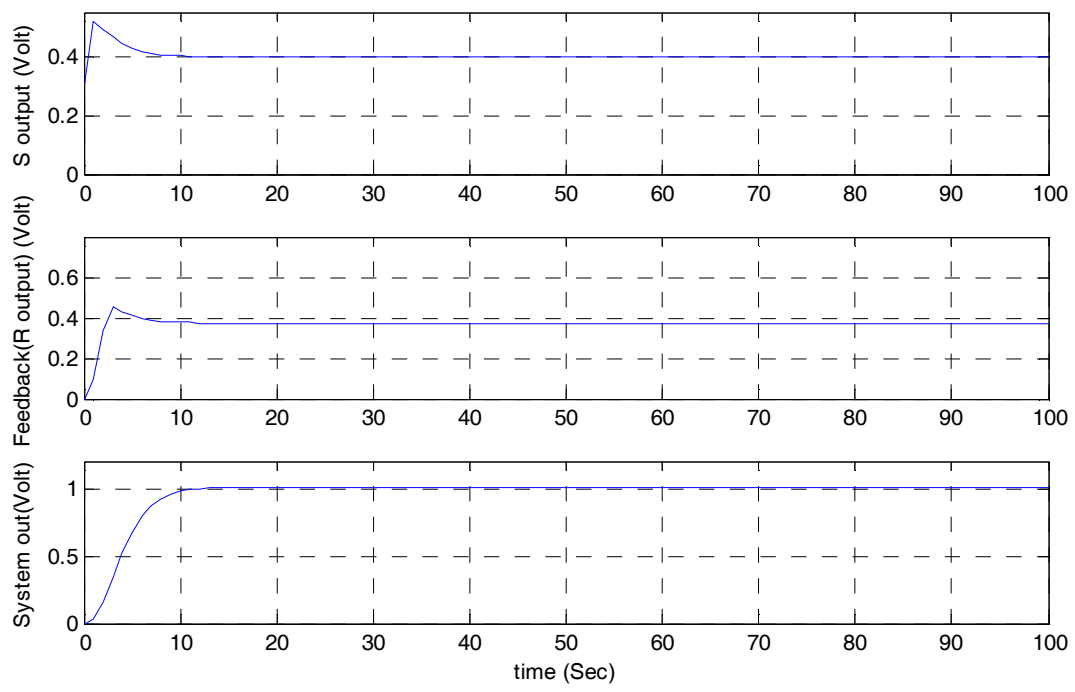

*Fig. 9 Matlab simulation of the RST controlled heat exchanger*

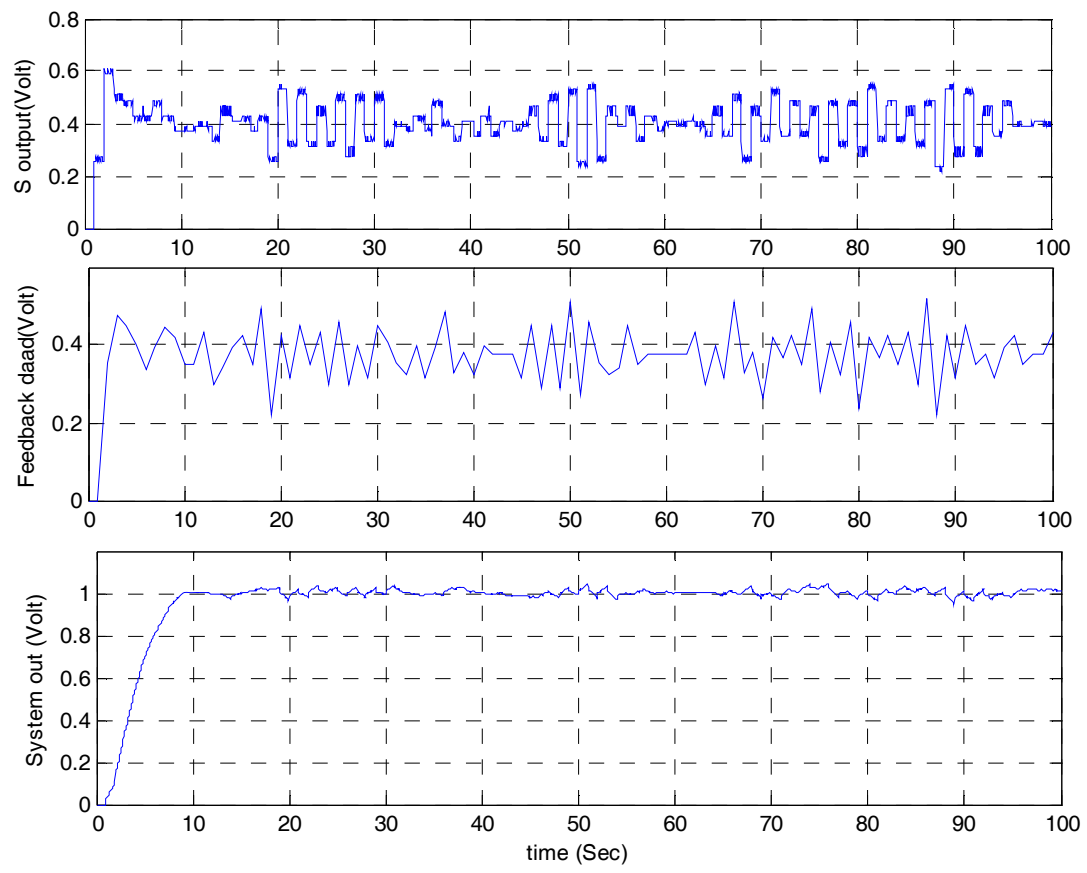

Fig. 10. The framework results for the RST controlled heat exchanger

The differences between fig 9 and 10 help ensures the fact that simulation is not equivalent to real time.

### *C. Hardware in the loop simulation for RST and heat exchanger*

The same RST controller with the same process is implemented in HiL simulation mode. It was implemented by C code on MC9S12E128 kit which includes ADC and DAC channels. They are used to interface the microcontroller with the framework. Fig 11 and Fig 12 show the framework implementation of the complete system in the HiL simulation mode, the system and control actions signals respectively.

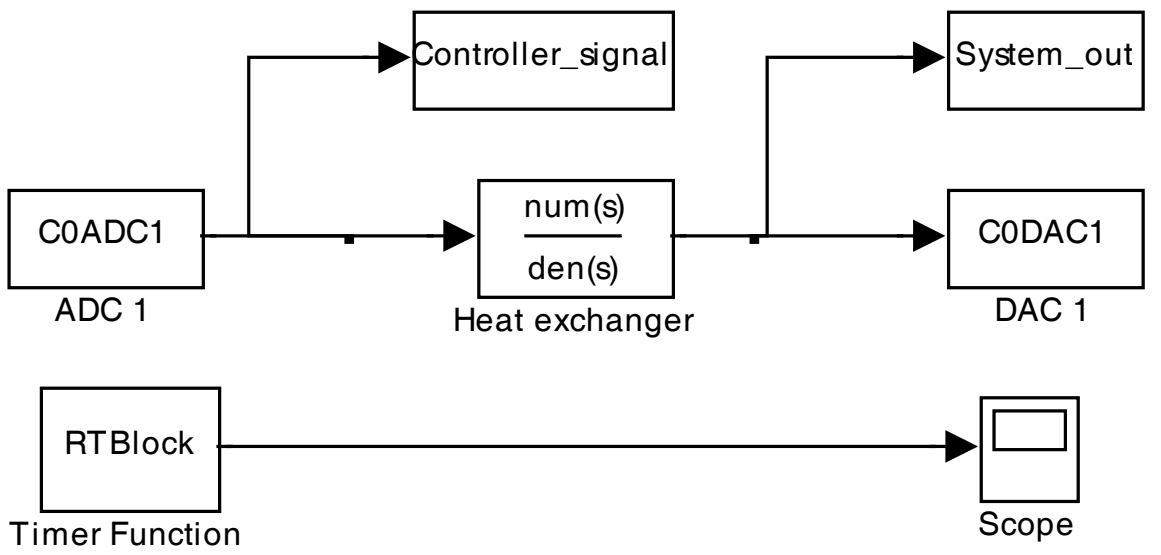

*Fig.11. Framework implementation of an RST Controller with a Heat Exchanger process in Hardware in the loop simulation*

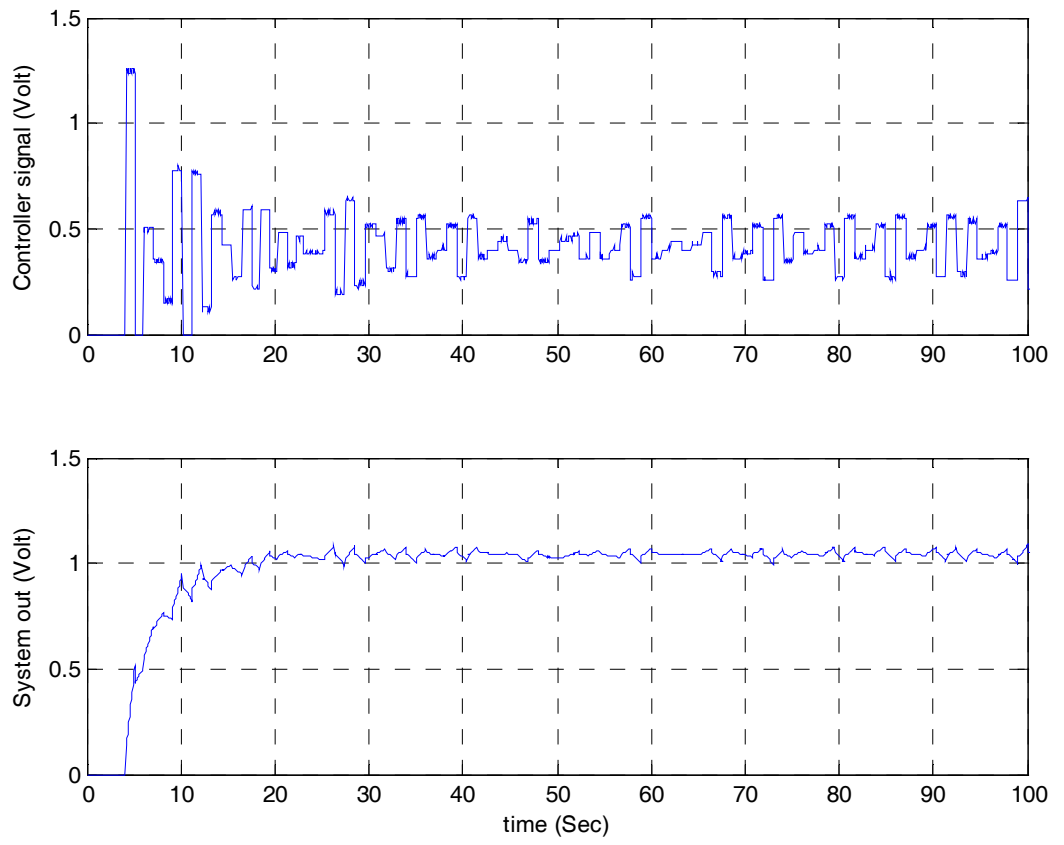

*Fig. 12 Results of the hardware in the loop simulation of the RST controlled heat exchanger*

In this mode the success of the control system is slightly difficult compared with the other mode. This is shown in the higher noise level associated with the output signal.

## Conclusion

This paper presented a framework that enable control engineers to use hardware in the loop simulation and real time simulations in control training and education. The framework used a low cost family of ADDA cards. The paper presented some software solutions for the nonlinearity, saturation and delays problems. Many basic and advanced control algorithms are tested and verified on the framework in both real time simulation and hardware in the loop simulation modes.